\documentclass[runningheads]{llncs}

\usepackage[utf8x]{inputenc}
\usepackage[T1]{fontenc}

\usepackage{amsmath}
\usepackage{amsfonts}
\usepackage{amssymb}
\usepackage{stmaryrd}

\usepackage{latexsym}
\usepackage{cmll}
\usepackage{graphicx}

\usepackage[dvipsnames]{xcolor}
\definecolor{aawhite}{rgb}{0.97,0.97,0.97}
\definecolor{awhite}{rgb}{0.90,0.90,0.90}
\definecolor{lgreen}{rgb}{0.94,1.0,0.98}
\definecolor{dgreen}{rgb}{0.0,0.3,0.1}
\definecolor{sgreen}{rgb}{0.0,0.7,0.3}
\definecolor{lgreen}{rgb}{0.94,1.0,0.98}
\definecolor{bgreen}{rgb}{0.00,0.50,0.25}
\definecolor{dblue}{rgb}{0.0,0.1,0.6}
\definecolor{lblue}{rgb}{0.8,0.8,1.0}
\definecolor{mixed}{rgb}{0.0,0.3,0.3}
\definecolor{dred}{rgb}{0.6,0.2,0.0}
\definecolor{sred}{rgb}{0.7,0.2,0.0}
\definecolor{ddred}{rgb}{0.3,0.1,0.0}
\definecolor{turq}{rgb}{0.28,0.82,0.80}
\definecolor{lyellow}{rgb}{1.00,0.97,0.94}
\definecolor{mygreen}{rgb}{0,0.6,0}
\definecolor{mygray}{rgb}{0.5,0.5,0.5}
\definecolor{mymauve}{rgb}{0.58,0,0.82}
\definecolor{codegreen}{rgb}{0,0.6,0}
\definecolor{codegray}{rgb}{0.5,0.5,0.5}
\definecolor{codepurple}{rgb}{0.58,0,0.82}
\definecolor{backcolour}{rgb}{0.96,0.96,0.96}

\usepackage{hyperref}
\hypersetup{
    colorlinks=true,
    linkcolor=sred,
    filecolor=magenta,
    urlcolor=cyan,
    citecolor=sgreen,
    bookmarks=true,
}

\usepackage{listings}
\renewcommand{\verb}{\lstinline}
\lstdefinelanguage{yarel} {
mathescape=true,
texcl=false,
keywords={iter, where, times, var, int, bool, module, import, dcl, def, dec, id, inc, neg, !dec, !id, !inc, !neg, if, then, else, elif, return, fix, !, it, rec},
morekeywords={package, public, class, static, void, new, throws, for, interface, private, final },
literate=	
	{0}{{\textcolor{orange}{0}}}1
	{1}{{\textcolor{orange}{1}}}1
	{2}{{\textcolor{orange}{2}}}1
	{3}{{\textcolor{orange}{3}}}1
	{4}{{\textcolor{orange}{4}}}1
	{5}{{\textcolor{orange}{5}}}1
	{6}{{\textcolor{orange}{6}}}1
	{7}{{\textcolor{orange}{7}}}1
	{8}{{\textcolor{orange}{8}}}1
	{9}{{\textcolor{orange}{9}}}1
	,
    morecomment=[l]{//}, 
    morecomment=[s]{/*}{*/}, 
    morestring=[s]{\\}{\\} 
}

\lstdefinestyle{yarel-style}{
	backgroundcolor=\color{backcolour},
	commentstyle=\color{codegreen},
	keywordstyle=\color{magenta},
	numberstyle=\tiny\color{codegray},
	stringstyle=\color{codepurple},
	basicstyle=\ttfamily\footnotesize,
	breakatwhitespace=false,
	breaklines=true,
	captionpos=b,
	keepspaces=true,
	numbers=left,
	numbersep=5pt,
	showspaces=false,
	showstringspaces=false,
	showtabs=false,
	tabsize=2
}
\lstdefinelanguage{java}{
    morekeywords = [1]{import, abstract, class, enum, extends
                      , implements, import, instanceof, interface, native
                      , new, final,  package, private, protected, public
                      , static, void},
    morekeywords = [2]{boolean, int, do, for, if, else, throws, catch, while
                      , try, null, length, assert, case, return, super, this},
    morekeywords = [3]{ },
    morekeywords = [4]{ },
    morekeywords = [5]{ },
    keywordstyle = [1]\color{magenta},
    keywordstyle = [2]\color{blue},
    keywordstyle = [3]\color{magenta},
    keywordstyle = [4]\color{orange},
    keywordstyle = [5]\color{lblue},
    sensitive = true,
    morecomment = [l]{//},
    morecomment = [s]{/*}{*/},
    morecomment = [s]{/**}{*/},
    commentstyle={\color{dgreen}},
    morestring = [b]",
    morestring = [b]',
    basicstyle={\small\ttfamily\bfseries},
    stringstyle={\ttfamily\small\color{orange}},
    numbers=left,
    numberstyle=\tiny\color{mygray},
    xrightmargin=0em,
    xleftmargin=3em,
    stepnumber=1,
    numbersep=1em,
    lineskip=-0.5ex,
    mathescape=true,
    showstringspaces=false,
    frame=none,
    breaklines=true,
    columns=[l]{fullflexible},
    keepspaces=true,
}

\lstnewenvironment{java}{\lstset{language={java},}}{}

\lstdefinestyle{java-style}{
	backgroundcolor=\color{backcolour},
	commentstyle=\color{codegreen},
	keywordstyle=\color{magenta},
	numberstyle=\tiny\color{codegray},
	stringstyle=\color{codepurple},
	basicstyle=\fontsize{9}{13}\selectfont\ttfamily,
	breakatwhitespace=false,
	breaklines=true,
	captionpos=b,
	keepspaces=true,
	numbers=left,
	numbersep=5pt,
	showspaces=false,
	showstringspaces=false,
	showtabs=false,
	tabsize=2
}
\newcommand{\vj}[1]{\lstinline[language=java,style=java-style]+#1+}

\begin{document}
\title{Splitting recursion schemes into reversible and classical interacting threads}

\author{Armando B. Matos\inst{1} \and
	Luca Paolini\inst{2}\orcidID{0000-0002-4126-0170} \and
	Luca Roversi\inst{2}\orcidID{0000-0002-1871-6109}}
\authorrunning{A. Matos, L. Paolini, L. Roversi}

\institute{Universidade do Porto, Departamento de Ciência de Computadores -- Portugal\\
	\email{armandobcm@yahoo.com}\\
	\and
	Università degli Studi di Torino, Dipartimento di Informatica -- Italy\\
	\email{\{luca.paolini,luca.roversi\}@unito.it}}

\maketitle
\begin{abstract}
Given a simple recursive function, we show how to extract from it a reversible and an classical iterative part. Those parts can synchronously cooperate under a Producer/Consumer pattern in order to implement the original recursive function. The reversible producer is meant to run on reversible hardware. We also discuss how to extend the extraction to a more general compilation scheme.
\end{abstract}

\section{Introduction}
\label{section:Introduction}

Our goal is to compile a class of recursive functions in a way that parts of the object code produced can leverage the promised green foot-print of truly reversible hardware. This work illustrates preliminary steps towards that goal. We focus on a basic class of recursive functions in order to demonstrate its feasibility.

\paragraph{Contributions.}
Let \vj{recF[p,b,h]}
be a recursive function
defined in some programming formalism, where
\vj{p} is a \emph{predecessor} function,
\vj{h} a \emph{step} function, and
\vj{b} a \emph{base} function.
We show how to compile \vj{recF[p,b,h]} into \vj{itFCls[b,h]} and \vj{itFRev[p,pInv]}
such that:
\begin{align}
\label{align:compilation scheme}
\mbox{\vj{recF[p,b,h]}} & \simeq
\mbox{\vj{itFCls[b,h]}} \parallel  \mbox{\vj{itFRev[p,pInv]}}
\enspace,
\end{align}
where:
(i) ``$ \simeq $'' stands for ``\emph{equivalent to}'';
(ii) \vj{itFCls[b,h]} is a classical
\vj{for}-loop that, starting from a value produced by \vj{b}, iteratively applies \vj{h};
(iii) \vj{itFRev[p,pInv]} is a reversible code with two \vj{for}-loops
in it one iterating \vj{p}, the other its inverse \vj{pInv};
(iv) ``$ \parallel $'' is interpreted as an \emph{interaction} between \vj{itFCls[b,h]} and \vj{itFRev[p,pInv]}, according to a Producer/Consumer pattern, where \vj{itFRev[p,pInv]} produces the values that \vj{itFCls[b,h]} consumes to implement the initially given recursion \vj{recF[p,b,h]}. In principle, \vj{itFRev[p,pInv]} can drive a real reversible hardware to exploit its low energy consumption features.

In this work we limit the compilation scheme~\eqref{align:compilation scheme} to use:
(i) a predecessor \vj{p} such that the value \vj{p(x)-x} is any \emph{constant} $ \Delta_{\mbox{\vj{p}}} $ equal to, or smaller than, \vj{-1};
(ii) recursion functions \vj{recF[p,b,h]} whose \emph{condition} identifying the base case is \vj{x<=0} instead than the more standard \vj{x==0}; this means that more than one base
\emph{non positive} value for \vj{recF[p,b,h]} exists
in the interval $ [ \Delta_{\mbox{\vj{p}}}+1, \mbox{\vj{0}} ] $.
This slight generalization will require a careful management of the
reversible behavior of \vj{itFRev[p,pInv]} and its interaction with \vj{itFCls[b,h]}
in order to reconstruct \vj{recF[p,b,h]}.

\paragraph{Contents.}
Section~\ref{section:The drivign idea} sets the stage to develop the main ideas about~\eqref{align:compilation scheme}, restricting \vj{recF[p,b,h]} to a recursive function that identifies its base case by means of the standard condition \vj{x==0};
this ease the description of how \vj{itFRev[p,pInv]} and \vj{itFCls[b,h]} interact.
Section~\ref{section:From recursion to iteration} extends \eqref{align:compilation scheme} to deal with \vj{recF[p,b,h]} having
\vj{x<=0}, and not \vj{x==0}, to identify its base case(s); this impacts on
how \vj{itFRev[p,pInv]} must work.
In both cases, the programming syntax we use can be interpreted
into the reversible languages \textsf{SRL}
\cite{DBLP:journals/tcs/Matos03,MatosRC2020} and \textsf{RPP} \cite{paolini2017ngc,DBLP:journals/tcs/PaoliniPR20,MatosRC2020},
up to minor syntactic details.
Section~\ref{section:Conclusions} addresses future work.
\begin{figure}
\begin{lstlisting}[
%float,
language = java,
style = java-style,
%caption = The recursive function \vj{recF}. ,
%label = primitive-recursive-function
]
 Fix recF(x)                          {
     if   (c(x)) { b(x);           }
     else        { h(x,recF(p(x))); } }
\end{lstlisting}
\caption{The recursive function \vj{recF}.}
\label{primitive-recursive-function}
\end{figure}

\begin{figure}
\begin{lstlisting}[
%float,
language = java,
style = java-style,
%caption = Iterative unfolding \vj{recF(3)}: the bottom-up part. ,
%label = iteratve simulation of f(3) recursive}
]
  /*** Assumption: the inital value of x is 3 */
  x = p(x)       // ==2
  x = p(x)       // ==1
  x = p(x)       // ==0
  y = b(x)       // ==b(p(p(p(3))))
  y = h(x,y)     // ==h(p(p(p(3))),b(p(p(p(3)))))
  x = pInv(x)    // ==pInv(p(p(p(3))))==p(p(3))
  y = h(x,y)     // ==h(p(p(3)),h(p(p(p(3))),b(p(p(p(3))))))
  x = pInv(x)    // ==pInv(p(p(3)))==p(3)
  y = h(x,y)     // ==h(p(3),h(p(p(3))
                 //         ,h(p(p(p(3))),b(p(p(p(3)))))))
  x = pInv(x)    // ==pInv(p(3))==3
  y = h(x,y)     // ==h(3,h(p(3),h(p(p(3))
                 //      ,h(p(p(p(3))),b(p(p(p(3))))))))
\end{lstlisting}
\caption{Iterative unfolding \vj{recF(3)}: the bottom-up part.}
\label{iteratve simulation of f(3) recursive}
\end{figure}

\section{The driving idea}
\label{section:The drivign idea}

Let \vj{recF[p,b,h]} in \eqref{align:compilation scheme} have a structure as in \textbf{Fig.\,\ref{primitive-recursive-function}} where
\vj{b(x)} is the \emph{base} function,
\vj{h(x,y)} the \emph{step} function,
\vj{p(x)} the \emph{predecessor} \vj{x-1}, and
\vj{c(x)} the \emph{condition} \vj{x==0} to identify a unique base case.
\par
\textbf{Fig.\,\ref{iteratve simulation of f(3) recursive}}
details out
\mbox{\vj{h(3,h(p(3),h(p(p(3)),}} \mbox{\vj{h(p(p(p(3))),b(p(p(p(3))))))))}},
unfolding  of \vj{recF(3)}.
Every comment asserts a property of the values that \vj{x} or \vj{y} stores.
Lines 2--4 unfold an iteration that computes \vj{p(p(p(3)))}, which eventually sets the value of \vj{x} to \vj{0}.
Line 5 starts the construction of the final value of \vj{recF(3)} by
applying the base case of \vj{recF}, i.e.~\vj{b(x)}.
By definition, let \vj{pInv} denote the inverse of \vj{p},
i.e.~\vj{pInv(p(z))==p(pInv(z))==z}, for any \vj{z}.
Clearly, in our running example, the function \vj{pInv(x)} is \vj{x\+1}.
Lines 6--13 alternate \vj{h(x,y)}, whose result \vj{y}, step by step, gets closer to the final value \vj{recF(3)}, and \vj{pInv(x)},
which produces a new value for \vj{x}.

\begin{figure}
\begin{lstlisting}[
%    float,
	language = java,
	style = java-style,
%	caption = Iterative \vj{itF} equivalent to \vj{recF}.,
%	label = iteratve version of recF
	]
	s = 0, e = 0, g = 0, w = 0
	w = w + x;
	for (i = 0; i<=w; i++)      {
		if      (x> 0) { g++; }
		else if (x==0) { e++; }
		else           { s++; }
		x = p(x);               }

	for (i = 0; i<=w; i++)                {
		x = pInv(x);
		if      (x> 0) { g--; y = h(x,y); }
		else if (x==0) { e--; y = b(x);   }
		else           { s--;             } }
	w = w - x;
\end{lstlisting}
\caption{Iterative \vj{itF} equivalent to \vj{recF}.}
\label{iteratve version of recF}
\end{figure}

Let us call \vj{itF} the code in \textbf{Fig.\,\ref{iteratve version of recF}}.
It implements \vj{recF} by means of finite iterations only.
Continuing with our running example, if we run \vj{itF} here above starting with
\vj{x==3}, then \vj{x==0} holds at line 8, just after the first \vj{for}-loop;
after the second \vj{for}-loop \vj{y==recF(3)} holds at line 14.

The code of \vj{itF} has two parts.
Through lines 2--7 the variable \vj{g} counts how many times \vj{x} remains positive, the variable \vj{e} how many it stays equal to \vj{0}, and the variable \vj{s} how many it becomes negative.
In this running example we notice that \vj{x} never
becomes negative, for the iteration at lines
3--7 is driven by the value of \vj{x} which, initially, we can assume
non negative, and which \vj{p(x)} decreases of a single unity.
We shall clarify the role of \vj{s} later.
Lines 9--13  undo what lines 2--7 do by executing
\vj{pInv(x)}, \vj{g--}, \vj{e--}, \vj{s--}, i.e.~the inverses, in reversed order, of
\vj{p(x)}, \vj{g\+\+}, \vj{e\+\+}, \vj{s\+\+}.
So the correct values of \vj{x} are available
at lines 12, and 11, ready to be used as arguments of
\vj{b(x)} and \vj{h(x,y)} to update \vj{y}
as in \textbf{Fig.\,\ref{iteratve version of recF}}, according to the results
we obtain by the recursive calls to \mbox{\vj{recF}}.

\begin{figure}
\begin{lstlisting}[
%float,
language = java,
   style = java-style,
%caption = Reversible side of \vj{itF}.,
%label = reversible core itF
]
 s = 0, e = 0, g = 0, w = 0
 w = w + x;
 for (i=0; i<=w; i++)      {
   if      (x> 0) { g++; } //number of times x is `g'reater than 0
   else if (x==0) { e++; } //number of times x is `e'qual to 0
   else           { s++; } //number of times x is `s'maller than 0
   x = p(x);               }

 for (i=0; i<=w; i++)                                           {
   x = pInv(x);
   if      (x> 0) { g--; /* Value of x for h availabe here */ }
   else if (x==0) { e--; /* Value of x for b availabe here */ }
   else           { s--;                                      } }
 w = w - x;
\end{lstlisting}
\caption{Reversible side of \vj{itF}.}
\label{reversible core itF}
\end{figure}

\par
Now, let us focus on the main difference between
\textbf{Fig.\,\ref{reversible core itF}} and
\textbf{Fig.\,\ref{iteratve version of recF}}.
\par
Both \vj{x=b(x)} and \vj{y=h(x,y)} at lines 12, and 11 of
\textbf{Fig.\,\ref{iteratve version of recF}}
are missing from lines 12, and 11 of \textbf{Fig.\,\ref{reversible core itF}}.
Dropping them let
\textbf{Fig.\,\ref{reversible core itF}} be the \emph{reversible side}
of \vj{itF}; calling \vj{b(x)} and \vj{h(x,y)} in it generates\vj{y},
which is the result we need, so preventing the possibility to reset
the value of every variable dealt with in \textbf{Fig.\,\ref{reversible core itF}}
to their initial value.
This is why we also need a \emph{classical side} of \vj{itF} that generates \vj{y} in collaboration with the \emph{reversible side} in order to implement the initial \vj{recF} correctly.

\begin{figure}
\begin{lstlisting}[
%float,
language = java,
style = java-style,
%caption = Classical side of \vj{itF}: the consumer \vj{itFCls}. ,
%label = loop on the classical side
]
  /*** Assumption. The value of the input x is available here */
  /* Inject the current x at line 2 of itFRev to let it start */
  iterations = /* Probe line 9 of itFRev to get the
                  number of iterations to execute   */
  y = b(/* Probe line 14 of itFRev to get the argument */);
  for (i = 0; i<iterations; i++)                  {
    y = h(/* Probe line 12 itFRev to get
             the first argument of h     */ , y); }
\end{lstlisting}
\caption{Classical side of \vj{itF}: the consumer \vj{itFCls}.}
\label{loop on the classical side}
\end{figure}

\begin{figure}
\begin{lstlisting}[
%float,
language = java,
style = java-style,
%caption = Reversible side of \vj{itF} updated to be the producer \vj{itFRev} of the values that the consumer \vj{itFCls} needs. ,
%label = loop on the reversible side
]
  s = 0, e = 0, g = 0, w = 0;
  x = /* Inject here the value of x at line 2 of itFCls */
  w = w + x;
  for (i = 0; i<=w; i++)      {
    if      (x> 0) { g++; }
    else if (x==0) { e++; }
    else           { s++; }
    x = p(x);                 }
  /* itFCls probes here g which has the number of iterations */
  for (i = 0; i<=w; i++)                                    {
    x = pInv(x);
    if      (x> 0) { g--; /* itFCls probes here the
                             first argument value of h */ }
    else if (x==0) { e--; /* itFCls probes here the
                             argument value of b       */ }
    else           { s--;                                 } }
  w = w - x;
\end{lstlisting}
\caption{Reversible side of \vj{itF} updated to be the producer \vj{itFRev} of the values that the consumer \vj{itFCls} needs.}
\label{loop on the reversible side}
\end{figure}

The previous observations lead to \textbf{Fig.\,\ref{loop on the classical side}}
which defines the \emph{classical side} \vj{itFCls} of \vj{recF},
and to \textbf{Fig.\,\ref{loop on the reversible side}} which defines the
\emph{reversible side} \vj{itFCRev} of \vj{recF}.

\medskip\par
So, here below we can illustrate how \vj{itFCls} and \vj{itFRev} synchronously interact, \vj{itFRev} producing values, \vj{itFCls} consuming them as arguments of \vj{b(x)} and \vj{h(x,y)}.

\medskip\par
Line 2 of \vj{itFCls} is the starting point of the synchronous interaction
between \vj{itFCls} and \vj{itFRev}; its comment:
\begin{align*}
&\mbox{\vj{/* Inject the current x at line 2 of itFRev to let it start */}}
   \end{align*}
describes what, in a fully implemented version of
\vj{itFCls}, we expect in that line of code.
The comment says that \vj{itFCls} injects (sends, puts)
its input value \vj{x} to line 2 of the
\emph{reversible side} \vj{itFRev} (cf. \textbf{Fig.\,\ref{loop on the reversible side}}).
Once \vj{itFRev} obtains that
value at line 2, as outlined by:
\begin{align*}
&\mbox{\vj{/* Inject here the value of x from line 2 of itFCls */}}
\end{align*}
\noindent
its \vj{for}-loop at lines 4--8 executes.
\par
After line 2, \vj{itFCls} stops at line 3. It waits for \vj{itFRev}
to produce the number of times that \vj{itFCls} has to iterate
line 7. Accordingly to:
\begin{align*}
	&\mbox{\vj{/*  Probe line 9 of itFRev to get the number of iterations to execute */}}
\end{align*}
\vj{itFRev} makes that value available in its variable \vj{g} at line 9:
\begin{align*}
	&\mbox{\vj{/* itFCls probes here g which has the number of iterations */}}
	\enspace .
\end{align*}
\par
Once gotten the value in \vj{iterations},
\vj{itFCls} proceeds to line 5 and
stops, waiting for \vj{itFRev} to produce the
argument of \vj{b} which is eventually available for
probing at line 14 of \vj{itFRev}.
\par
Once the argument becomes available \vj{b} is applied,
and \vj{itFCls} enters its \vj{for}-loop, stopping at line 7 at every iteration.
The reason is that \vj{itFCls} waits for line 12 in \vj{itFRev} to produce the value of
the first argument of \vj{h(x,y)}. This interleaved dialog between line 7 of \vj{itFCls}
and line 12 of \vj{itFRev} lasts \vj{iterations} times.
\begin{figure}
\begin{lstlisting}[
%float,
language = java,
style = java-style,
%caption = The generic structure of \vj{recG}.  ,
%label = recG
]
  Fix recG(x)                      {
    if (x<=0) { b(x);            }
    else      { h(x,recG(p(x))); } }
\end{lstlisting}
\caption{The generic structure of \vj{recG}.}
\label{recG}
\end{figure}

\section{From recursion to iteration}
\label{section:From recursion to iteration}
We now generalize what we have seen in Section~\ref{section:The drivign idea}.
Inside \eqref{align:compilation scheme} we use \vj{recG} of \textbf{Fig.\,\ref{recG}}
instead than \vj{recF} of \textbf{Fig.\,\ref{primitive-recursive-function}}.
This requires to generalize \textbf{Fig.\,\ref{loop on the reversible side}}.

\par
From the introduction we recall that, given a \emph{predecessor}
\vj{p(x)}, we define
$ \Delta_{\mbox{\vj{p}}} = \mbox{\vj{p(x)-x}}$, which is
a negative value.
In this section $ \Delta_{\mbox{\vj{p}}} $ can be any \emph{constant} $ k\,\mbox{\vj{<=-1}} $, not only
$ k\,\mbox{\vj{==-1}} $; this
requires to consider the slightly more general
\emph{condition} \vj{x<=0} in \vj{recG}.
For example, let \vj{p(x)} be \vj{x-2}.
The computation of \vj{recG(3)} is
\vj{h(3,h(p(3),h(p(p(3)),b(p(p(3))))))} which looks for the least
$ n $  of iterated applications of \vj{p(x)} such that
\vj{p(...p(3)...)<=0}; in our case we have $\mbox{\vj{2==}}\, n\, \mbox{\vj{<3}} $.

\begin{figure}
\begin{lstlisting}[
%float,
language = java,
style = java-style,
%caption = The iterative reversible function \vj{itG}.,
%label = iteratve recG
]
  s = 0, e = 0, g = 0, w = 0;
  z = 0, predDivX = 0, predNotDivX = 1;
  w = w + x; /* x is assumed to be the input */
  for (i = 0; i <= w; i++)   {
    if      (x >  0) { g++; }
    else if (x == 0) { e++; }
    else             { s++; }
    x = p(x);                  }

  for (i = 0; i < e; i++)                 {
    predDivX = predDivX + predNotDivX;
    predNotDivX = predDivX - predNotDivX; }

  for (j = 0; j < predDivX; j++)            {
  	for (i = 0; i <= w; i++)               {
  		x = pInv(x);
  		if      (x >  0) { g--; y = h(x,y); }
  		else if (x == 0) { e--; y = b(x);   }
  		else             { s--;             }}}

  for (j = 0; j < predNotDivX; j++)                           {
    w++;
    for (i = 0; i <= w; i++)                                 {
    	x = pInv(x);
    	if      (x >  0) { g--;
                         x = p(x);
                         if      (z <  0) {                }
                         else if (z == 0) { y = b(x); z++; }
                         else             { y = h(x,y);    }
                         x = pInv(x);                       }
    	else if (x == 0) { e--;             }
    	else             { s--;             }}
    w--;                                                      }
  for (i = 0; i < predNotDivX; i++)  {
    z--;                             }
  w = w - x;
  /* y carries the output */
\end{lstlisting}
\caption{The iterative function \vj{itG}.}
\label{iteratve recG}
\end{figure}

\medskip\par
\textbf{Fig.\,\ref{iteratve recG}} introduces \vj{itG}
which generalizes \vj{itF} in \textbf{Fig.\,\ref{iteratve version of recF}}.

\medskip\par
The scheme \vj{itG} iteratively implements any recursive function whose
structure can be brought back to the one of \vj{recG}.
We remark that line 1 in
\textbf{Fig.\,\ref{iteratve recG}} initializes ancillae
\vj{s}, \vj{e}, \vj{g}, and \vj{w}, like
\textbf{Fig.\,\ref{iteratve version of recF}}
initializes the namesake variables of \vj{itF},
but line 2 of \vj{itG} has new ancillae \vj{z}, \vj{predDivX}, and \vj{predNotDivX}.

\par
We also assume an initial \emph{non negative} value for \vj{x}.
The reason is twofold.
Firstly, it keeps our discussion as simple as possible,
with no need to use the absolute value of \vj{x} to set
the upper limit of every index \vj{i}
in the \vj{for}-loops that occur in the code.
Second, negative values of \vj{x} would widen our discussion about
what a classical recursive function on negative values is and about
what its reversible equivalent iteration has to be; we see this as a
very interesting subject connected to \cite{BOITEN1992139}, which is much more oriented than us to optimization issues of recursively defined functions.

\par
We start observing that line 3 of \vj{itG} sets \vj{w} to the initial value
of \vj{x}; the reason is that every \vj{for}-loop, but the one at lines 10--12, has to last \vj{x\+1} iterations, and \vj{x} changes in the course of the computation; so, \vj{w} stores the initial value of \vj{x}
and stays constant from line 4 through line 21.
In fact it can change at lines 22--33. We will see why, but
\vj{w} is eventually reset to its initial value \vj{0} at line 36.

With the here above assumptions, given a non negative \vj{x},
and in analogy to \vj{itF}, the \vj{for}-loop at lines 4--8 of \vj{itG}
iterates the application of \vj{p(x)} as many times as \vj{w\+1},
i.e.~the initial value of \vj{x} plus \vj{1}.
So, the value of \vj{x} at line 9 is equal to
$ \mbox{\vj{w\+(w\+1)*}}\Delta_{\mbox{\vj{p}}} $ which cannot be positive.
In particular, all the values that \vj{x} assumes in the \vj{for}-loop at
lines 4--8 belong to the following interval:
\begin{align}
	\label{align:interval of the iteration}
	I(\mbox{\vj{w}})
	& \triangleq [
	\mbox{\vj{w\+(w\+1)*}}\Delta_{\mbox{\vj{p}}}
	,\mbox{\vj{w\+w*}}\Delta_{\mbox{\vj{p}}}
	,\ldots
	,\mbox{\vj{w\+}}\Delta_{\mbox{\vj{p}}}
	,\mbox{\vj{w}}]
\end{align}
from the least to the greatest;
the counters \vj{g}, \vj{e}, \vj{s} say how many elements of
$ I(\mbox{\vj{x}}) $ are \vj{g}reater, \vj{e}qual or \vj{s}maller
than \vj{0}, respectively.
Depending on \vj{0} to belong to $ I(\mbox{\vj{x}}) $ determines the
behavior of the reminder part of \vj{itG}, i.e.~lines 10--36.

\par\vspace{\baselineskip}
We distinguish two cases in order to illustrate them.

\paragraph{First case.}
Let $ \mbox{\vj{w\%}}\Delta_{\mbox{\vj{p}}}\ \mbox{\vj{== 0}} $,
i.e.~the integer value $ \Delta_{\mbox{\vj{p}}} $ divides with no reminder the initial value of \vj{x} that we find in \vj{w}.
So, $ \mbox{\vj{0}} \in I(\mbox{\vj{x}}) $, which implies the following
relations hold at line 9:
\begin{align}
	\mbox{\vj{e == 1}}
	&&
	\mbox{\vj{g == -}} \frac{\mbox{\vj{w}}}
	                        {\Delta_{\mbox{\vj{p}}}}
	&&
	\mbox{\vj{s == (w\+1)-g-e}}
	\enspace.
\end{align}

\begin{figure}
\begin{lstlisting}[
language = java,
style = java-style,
%caption = A possible replacement of lines 10--12 in \textbf{Listing~\ref{iteratve recG}}.,
%label = if alternative to for
]
 if      (e <  0) {                                       }
 else if (e == 0) { predDivX = predDivX+predNotDivX;
                    predNotDivX = predDivX - predNotDivX; }
 else             {                                       }
\end{lstlisting}
\caption{A possible replacement of lines 10--12 in \textbf{Fig.\,\ref{iteratve recG}}.}
\label{if alternative to for}
\end{figure}

Lines 10--12 execute exactly once, swapping \vj{predDivX} and \vj{predNotDivX}.
As a remark, we could have well used the
\vj{if}-selection in \textbf{Fig.\,\ref{if alternative to for}}
(a construct of \textsf{RPP}) in place of the \vj{for}-loop at lines 10--12, but we opt for a more compact code.

\par
Swapping \vj{predDivX} and \vj{predNotDivX} sets \vj{predDivX==1}
and \vj{predNotDivX==0}, computationally exploiting that
$ \Delta_{\mbox{\vj{p}}} $ divides \vj{w} with no reminder:
the \vj{for}-loop body
at lines 15--19 becomes accessible, while lines 22--33, with
\vj{for}-loops among them, do not. Lines 15--19 are identical to
lines 10--16 of \vj{itF} in
\textbf{Fig.\,\ref{reversible core itF}} which
we already know to correctly apply \vj{b(x)} and \vj{h(x,y)} in
order to simulate the recursive function we start from.

\paragraph{As a second case.}
Let $ \mbox{\vj{w\%}}\Delta_{\mbox{\vj{p}}}\, \mbox{\vj{!= 0}} $,
i.e.~the integer value $ \Delta_{\mbox{\vj{p}}} $ divides the initial value of \vj{x} that \vj{w} stores, \emph{but with some reminder}.
So, $ \mbox{\vj{0}} \not\in I(\mbox{\vj{x}}) $, which imply:
\begin{align}
\label{align:egs non divide}
	\mbox{\vj{e == 0}}
	&&
	\mbox{\vj{g == -}}
	\left\lfloor\frac{\mbox{\vj{w}}}{\Delta_{\mbox{\vj{p}}}}\right\rfloor
	&&
	\mbox{\vj{s == (w\+1)-g-e}}
\end{align}
\noindent
hold at line 9. Lines 11--12 cannot execute, leaving \vj{predDivX} and \vj{predNotDivX} as they are: lines 22--33 become accessible and the \vj{for}-loop at lines 15--19 does not.
\par\noindent
Line 22 increments \vj{w} to balance the information loss that the
rounding of \vj{g} in \eqref{align:egs non divide} introduces;
line 33 recovers the value of \vj{w} when the outer \vj{for}-loop
starts.

\par\noindent
The \vj{if}-selection at lines 25--32 identifies when
to apply \vj{b(x)}, which must be followed by the required applications
of \vj{h(x,y)}. We know that $ \mbox{\vj{0}} \not\in I(\mbox{\vj{x}}) $,
so \vj{x==0} can never hold.
Clearly, \vj{s--} is executed until \vj{x>0}.
But the \emph{first} time \vj{x>0} holds true we must compute \vj{b(p(x))},
because the \emph{base} function \vj{b(x)} \emph{must be used the last time} \vj{x} assumes a negative value, \emph{not the first time} it gets positive;
lines 26--30 implement our needs.
Whenever \vj{x>0} is true, the value of \vj{x} is one step ahead the
required one: we get one step back with line 26 and,
if it is the first time we step back, i.e.~\vj{z==0} holds,
then we must execute line 28. If not, i.e.~\vj{z!=0},
we must apply the  \emph{step} function at line 29.
Line 30, restores the right value of \vj{x}.
Finally, the \vj{for}-loop at line 34 sets \vj{z} to its initial value.

\medskip
At this point, in order to obtain the fully reversible version of \textbf{Fig.\,\ref{iteratve recG}} we must think of replacing
the calls to \vj{h(x,y)} and \vj{b(x)} at lines in 28, and 29
by means of actions that probe the value of \vj{x}, in analogy to
\textbf{Fig.\,\ref{loop on the reversible side}}, lines 12 and 14. The full details are in \cite{Javacode} which we look as a playground with \textsf{Java} classes that implement \textbf{Fig.\,\ref{iteratve recG}} and \textbf{Fig.\,\ref{loop on the classical side}}
as synchronous and parallel threads, acting as a producer and a consumer.
\section{Future work}
\label{section:Conclusions}
We have shown that we can decompose every classical recursive function, based on a \emph{predecessor} that decreases every of its input by a constant value, into reversible and classical components that cooperate to implement the original recursive functions under a Producer/Consumer pattern (see \eqref{align:compilation scheme}).

Firstly, we plan to extend \eqref{align:compilation scheme} to recursive functions \vj{recF} based on predecessors \vj{p} not limited to a constant $ \Delta_{\mbox{\vj{p}}} $ not greater than \vj{-1}. A predecessor \vj{p} should be at least such that:
\begin{enumerate}
	\item
	$ \Delta_{\mbox{\vj{p}}} $ is not necessarily a constant.
    For example, $ \Delta_{\mbox{\vj{p}}}\ \mbox{\vj{== -3}}$ on even arguments, and \vj{-2} on odd ones can be useful;
	\item
	the predecessor can be an integer division \vj{x/k}, for some given \vj{k>0}, like in a dichotomic search, which has
    \vj{k==2}.
\end{enumerate}
Secondly, we aim at generalizing \eqref{align:compilation scheme} to a compiler $ \llbracket \cdot \rrbracket $:
\begin{equation}
\label{equation:compilation scheme detailed}
\begin{split}
	\llbracket \mbox{\vj{p}} \rrbracket
    & = \mbox{\vj{some implementation code}}
	\\
	\llbracket \mbox{\vj{pInv}} \rrbracket & =
   \mbox{\vj{!}}\llbracket \mbox{\vj{p}} \rrbracket
    \mbox{\vj{, i.e. implementation that inverts}}\
    \llbracket \mbox{\vj{p}} \rrbracket
    \\
	\llbracket \mbox{\vj{recF[p,b,h]}} \rrbracket & =
	\mbox{\vj{itFCls[$\llbracket$b$\rrbracket$,$\llbracket$h$\rrbracket$]}}
	\parallel \mbox{\vj{itFRev[$\llbracket$p$\rrbracket$,$\llbracket$pInv$\rrbracket$]}}
    \enspace .
\end{split}
\end{equation}
The domain of $ \llbracket \cdot \rrbracket $ should be a class \textsf{R} of recursive functions
built by means of standard composition schemes,
starting from a class of predecessors \vj{p1}, \vj{p2}, \ldots each of which must have the corresponding inverse function \vj{p1Inv}, \vj{p2Inv}, \ldots.

In these lines we want to explore interpretations of $ || $  more liberal than the essentially obvious synchronous Producer/Consumer that we implement in \cite{Javacode}.
We shall very likely take advantage of parallel discrete events simulators as described in \cite{SchordanOJB18,Schordan2020} in order to get rid of any explicit synchronization between the pairs of reversible-producer/classical-consumer that
\eqref{equation:compilation scheme detailed} would recursively generate when applied to an element in \textsf{R}.

\medskip
We also plan to follow a more abstract line of research.
The compilation scheme \eqref{equation:compilation scheme detailed}  recalls Girard's decomposition
$ A \rightarrow B \simeq\ !A \multimap B $ of a classical computation into a linear one that can erase/duplicate computational resources.
Decomposing \vj{recF[p,b,h]} in terms of
\vj{itFCls[b,h]} and \vj{itFRev[p,pInv]} suggests that the relation between
reversible and classical computations can be formalized by
a linear isomorphism $ A^n \multimapboth B^n$ between tensor products
$ A^n$, and $B^n $ of $ A$, and $B $, in analogy to  \cite{DBLP:conf/popl/JamesS12}. Then we can think of recovering classical computations by some functor, say $ \gamma $, whose purpose is, at least, to forget, or to inject replicas, of parts
of $ A^n$, and $B^n $ in a way that
$ (\gamma A^n \rightarrow \gamma A^n) \uplus
  (\gamma A^n \leftarrow  \gamma A^n)  $
can be their type.
The type says that we move from a reversible computation to a classical one
by choosing which is input and which is output, so recovering the freedom
to manage computational resources as we are used to when writing classical programs.

\bibliographystyle{abbrv}
\bibliography{bib-minimal}

\appendix
\section{Reversible Producer/Irreversible Consumer in \textsf{Java}}
\label{section:Iteration as Producer/Consumer}
We briefly comment the \textsf{Java} code in \cite{Javacode} which compiles and runs with
\href{https://www.oracle.com/Java/technologies/Javase/jdk14-archive-downloads.html}{\textsf{Java SE 14 - ORACLE}}.

The section is meant to be self-contained; having an intuitive idea about what a \textsf{Java} thread is should be more than sufficient
to catch the essential points. Endless literature on \textsf{Java} is of public domain and \cite{Javacode} is based on it.

\par
\textsf{Java} classes in \cite{Javacode} are based on threads which,
synchronously collaborating in accordance with a Producer/Consumer template
by means of a couple of communication channels, \emph{implement}
both \vj{itFCls} in \textbf{Fig.\,\ref{loop on the classical side}},
and \vj{itG} in \textbf{Fig.\,\ref{iteratve recG}}
corresponding to some recursive function traceable to \vj{recG} in \textbf{Fig.\,\ref{recG}}.

\begin{figure}
\begin{lstlisting}[
%float,
language = Java,
style = Java-style,
%caption = The (irreversible) consumer class \vj{ItGCls}.,
%label = ItGCls
]
  public class ItGCls                                  {
    private final Inject inject;
    private final Probe probe;
    private int out = 0;
    private int in = 0;
    public ItGCls(Inject inject, Probe probe, int x) {
      this.inject = inject;
      this.probe = probe;
      this.in = x;                                   }
    public int getOut() { // Let out be available outside ItGCls
      return this.out;  }
    public void itFCls() throws InterruptedException {
      inject.put(in);
      int iterations = probe.get();
      out = B.b(probe.get());
      for (int i = 0; i < iterations; i++) {
        out = H.h(probe.get(), out);     }       }    }
\end{lstlisting}
\caption{The (irreversible) consumer class \vj{ItGCls}.}
\label{ItGCls}
\end{figure}

\par
The classes are \vj{ItGCls} (\textbf{Fig.\,\ref{ItGCls}}),
and \vj{ItGRev} (\textbf{Fig.\,\ref{ItGRev}}).
\par
They define methods
\vj{ItGCls.itGCls} and \vj{ItGRev.itGRev}, respectively.
In particular, \vj{ItGRev.itGRev} implements \vj{itG} in
\textbf{Fig.\,\ref{iteratve recG}}.

As a global assumption, \vj{B} and \vj{H} are two
\textsf{Java} classes with (\vj{static}) methods \vj{b(x)} and \vj{h(x,y)}, respectively,
one implementing a \emph{base} function, the other a \emph{step}
function. For example, downloading them from \cite{Javacode},
one finds a \vj{B.b(x)} which is the identity, and
a \vj{H.h(x,y)} that returns \vj{x\+y}.

As outlined, \vj{ItGCls} in \textbf{Fig.\,\ref{ItGCls}} details out the classical side of \vj{itG}, coherently with \textbf{Fig.\,\ref{loop on the classical side}}.
Lines 15--17 of \vj{ItGCls} are a classical iteration which,
given a number of \vj{iterations}, and the right sequence of values in the
argument variable \vj{in}, executes the sequence of assignments:
\begin{eqnarray}
\nonumber
&\mbox{\vj{out=B.b(probe.get())}}\\
\label{align:sequence assignments}
&\mbox{\vj{out=H.h(probe.get(),out);}}\\
\nonumber
&\mbox{\vj{out=H.h(probe.get(),out);}}\\
\nonumber
&\mbox{\vj{...}}
\end{eqnarray}
until \vj{out} contains the result, i.e.~what we call \vj{y} in \textbf{Fig.\,\ref{loop on the classical side}}.
At lines 7--9, \vj{ItGCls} sets the instances \vj{inject} and \vj{probe}
of the two communication channels \vj{Probe} and \vj{Inject}.
Lines 15, and 17 obtain the required argument values of the
\emph{base} and of the \emph{step} functions by calling \vj{probe.get()}.
Also line 14 calls \vj{probe.get()}. It is to obtain the number of \vj{iterations}. Instead, line 13 sends the initial value of \vj{in}
to the reversible side, i.e.~(an instance of) the producer \vj{ItGRev};
\vj{ItGRev} needs such a value
to start producing the sequence of arguments values that \vj{ItGCls}
requires to execute the sequence~\eqref{align:sequence assignments}.

\begin{figure}
\begin{lstlisting}[
%float,
language = Java,
style = Java-style,
%caption = The reversible producer \vj{ItGRev}.,
%label = ItGRev
]
 public class ItGRev                                      {
   private final Inject inject;
   private final Probe probe;
   ItGRev(Inject inject, Probe probe) {
     this.inject = inject;
     this.probe = probe;            }
   public void itGRev() throws InterruptedException      {
     int s = 0, e = 0, g = 0, w = 0, x = 0;
     int predDivX = 0, predNotDivX = 1;
     x = inject.swapIn(x); // read x from itGCls
     w = w + x;
     for (int i = 0; i <= w; i++)  {
       if      (x >  0) { g++; }
       else if (x == 0) { e++; }
       else             { s++; }
       x = Pred.pred(x);           }
     for (int i = 0; i < e; i++)               {
       predDivX = predDivX + predNotDivX;
       predNotDivX = predDivX - predNotDivX;   }
     for (int j = 0; j < predDivX; j++)      {
       probe.put(g); // send g to itGCls for iterations
       for (int i = 0; i <= w; i++)         {
         x = Pred.predInv(x);
         if      (x >  0) { g--;
                            // send x to itGCls for h(x,y)
                            probe.put(x); }
         else if (x == 0) { e--;
                            // send x to itGCls for b(x)
                            probe.put(x); }
         else             { s--;          }}}
     for (int j = 0; j < predNotDivX; j++)         {
       probe.put(g); // send g to itGCls for iterations
       w++;
       for (int i = 0; i <= w; i++)              {
         x = Pred.predInv(x);
         if      (x >  0) { g--;
                            x = Pred.pred(x);
                            // send x to itGCls for b(x) or h(x,y)
                            probe.put(x);
                            x = Pred.predInv(x);}
         else if (x == 0) { e--;                }
         else             { s--;                }}
       w--;                                       }
     w = w - x;
     x = inject.swapOut(x); // restore initial x  }}
\end{lstlisting}
\caption{The reversible producer \vj{ItGRev}.}
\label{ItGRev}
\end{figure}

As told, \textbf{Fig.\,\ref{ItGRev}} is a possible implementation of \textbf{Fig.\,\ref{iteratve recG}}.
The table:
\begin{center}
\begin{tabular}{c|c}
 \mbox{\vj{itG}}     & \mbox{\vj{ItGRev}} \\\hline\hline
 4--8 \quad   & 12--16      \\\hline
10--12\quad   & 17--19      \\\hline
14--19\quad   & 20--30      \\\hline
21--33\quad   & 31--43
\end{tabular}
\end{center}
lists code lines in \vj{itG}
of \textbf{Fig.\,\ref{iteratve recG}} that correspond to lines in
\vj{ItGRev} of \textbf{Fig.\,\ref{ItGRev}}.

We now trace the behavior of \vj{ItGRev} in analogy to the one of \vj{itG}.
Let $ \mbox{\vj{w\%}}\Delta_{\mbox{\vj{p}}}\mbox{\vj{== 0}} $, so lines 21--30 are accessible.
Line 21 sets the value of \vj{iterations} in
the consumer \vj{ItGCls}, so \vj{ItGCls} can move to line 15
waiting for the argument value of \vj{B.b(x)}.
Lines 26, and 29 call \vj{probe.put()}; in both cases
the call sets the value that the consumer
\vj{ItGCls} waits in order to feed \vj{H.h(x,out)} or \vj{B.b(x)}.

Let, instead, $ \mbox{\vj{w\%}}\Delta_{\mbox{\vj{p}}}\mbox{\vj{!= 0}} $,
so lines 32--43 are accessible.
Line 32 sets the value of \vj{iterations} in
the consumer \vj{ItGCls} exactly like line 21.
Lines 36--42 are a simplification of lines 25--32 of \vj{itG} in
\textbf{Fig.\,\ref{iteratve recG}}, for \emph{they do not need}
neither an \vj{if}-selection nor \vj{z} to identify which between
\vj{B.b(x)} and \vj{H.h(x,out)} to use, operation delegated to \vj{ItGCls}
which receives \vj{x} from \vj{probe.put(x)} at line 39.
This is also why lines 34--35 in \textbf{Fig.\,\ref{iteratve recG}} have no correspondent in \textbf{Fig.\,\ref{ItGRev}}.

Finally, the producer \vj{ItGRev} is triggered to start at line 10:
\vj{inject.swapIn()} swaps the content of \vj{x}, ancilla local to
\vj{ItGRev}, with the input value, (possibly) available in \vj{Inject.inject},
shared between \vj{ItFRev} and \vj{ItGCls}.
Line 45 restores the initial value of \vj{x} by swapping it with the one
it gets at line 10.

\begin{figure}
\begin{lstlisting}[
%float,
language = Java,
style = Java-style,
%caption = Channel \vj{Probe}.,
%label = Probe
]
  public class Probe                                             {
  	private int x = 0;
  	private boolean xAvailable = false;
  	public synchronized int get()
  	         throws InterruptedException                {
  		while (!xAvailable) { wait(); } // producer has not produced
  		int out = x;
  		xAvailable = !xAvailable;
  		notify();
  		return out;                                             }
  	public synchronized void put(int in)
  	         throws InterruptedException                   {
  		while (xAvailable) { wait(); } // consumer has not consumed
  		x = in;
  		xAvailable = !xAvailable;
  		notify();                                                 }}
\end{lstlisting}
\caption{Channel \vj{Probe}.}
\label{Probe}
\end{figure}

\begin{figure}
\begin{lstlisting}[
%float,
language = Java,
style = Java-style,
%caption = Channel \vj{Inject}.,
%label = Inject
]
  public class Inject                         {
    private int xInitial = 0;
    private boolean notSet = true;
    public synchronized int get()
            throws InterruptedException      {
      while (notSet) { wait(); }
      int out = xInitial;
      notify();
      return out;                            }
    public synchronized void put(int in)
             throws InterruptedException     {
      while (!notSet) { wait(); }
      xInitial = in;
      notSet = !notSet;
      notify();                              }
    public synchronized int swapIn(int in)
            throws InterruptedException      {
      while (notSet) { wait(); }
      int out = xInitial;
      xInitial = in;
      notify();
      return out;                            }
    public synchronized int swapOut(int in)
             throws InterruptedException     {
      int out = xInitial;
      xInitial = in;
      notSet = !notSet;
      notify();
      return out;                            }}
\end{lstlisting}
\caption{Channel \vj{Inject}.}
\label{Inject}
\end{figure}

Instances of classes \vj{Probe} and \vj{Inject} model channels
for synchronous dialogues between (instances of) \vj{ItGCls} and \vj{ItGRev}.
\textsf{Java} `\vj{synchronized}' directive let the bodies
of every method in \vj{Probe} and \vj{Inject} be critical regions,
each one accessed by a single thread at a time, one running \vj{ItGCls}, the other \vj{ItGRev}.
Let us comment on method \vj{get} in \vj{Probe}, all others, those ones of
\vj{Inject} included, behaving analogously
\footnote{We address to the official documentation on \textsf{Java} classes \vj{Runnable} and \vj{Threads} on possible implementations of critical regions.}.
Let \vj{probe.get()} be a call in \vj{ItGCls} of an instance \vj{probe} of \vj{Probe}.
The consumer enters \vj{probe.get()} and \vj{waits()} until
\vj{xAvailable} is set to \vj{true}, meaning that the producer \vj{ItGRev}
made a value available in \vj{probe.x} by means of a call
to \vj{probe.put(x)}. As soon as \vj{ItGRev} negates \vj{xAvailable},
setting it to \vj{true}, and calls \vj{notify()} at line 16 in
\vj{probe.put(x)}, \vj{ItGCls} leaves line 6 of \vj{probe.get()} in order to:
(i) set \vj{out} with the value in \vj{x};
(ii) negate \vj{xAvailable} again;
(iii) notify that \vj{ItGRev} can produce a new value for \vj{probe.x};
(iv) return the value of \vj{probe.out}.
Method \vj{prob.put(x)} sets a symmetric behavior on \vj{ItGRev}.
\end{document}